\documentclass[aps,twocolumn,prb,showpacs,eqsecnum]{revtex4}
\usepackage{graphicx}
\usepackage{amsfonts}
\usepackage{amsmath}
\usepackage[figuresright]{rotating}  
\usepackage{amssymb}
\usepackage{amsmath}
\usepackage{psfrag}
\usepackage{subfigure}
\usepackage{multirow}
\usepackage{tabularx}
\usepackage{textcomp}
\usepackage{units}
\usepackage{mathrsfs}
\usepackage{bm}

\def\nn{\nonumber}

\def\beq{\begin{eqnarray}}
\def\eeq{\end{eqnarray}}
\def\c{\hspace{2pt}}
\def\up{\uparrow}
\def\down{\downarrow}
\def\G{\mathcal{G}}

\renewcommand{\v}[1]{\ensuremath{\mathbf{#1}}} 
 
\newcommand{\uv}[1]{\ensuremath{\mathbf{\hat{#1}}}} 
\let\baraccent=\= 
\renewcommand{\=}[1]{\stackrel{#1}{=}} 

\begin{document}

\title{Unconventional magnetism in imbalanced Fermi systems with magnetic dipolar interactions}
\author{Benjamin M. Fregoso and Eduardo Fradkin}
\affiliation{Department of Physics, University of Illinois, 1110 West Green Street, Urbana, Illinois 61801-3080, USA}

\begin{abstract}
We study the magnetic structure of the ground state of an itinerant Fermi system of spin-\nicefrac{1}{2} particles with magnetic dipole-dipole interactions. We show that, quite generally, the spin state of particles depend on its momentum, i.e.,  spin and orbital degrees of freedom are entangled and taken separately are not ``good'' quantum numbers. Specifically, we consider a uniform system with non-zero magnetization at zero temperature. Assuming the magnetization is along $z$-axis, the quantum spin states are $\v{k}$-dependent linear combinations of eigenstates of the  $\sigma_z$ Pauli matrix. This leads to novel spin structures in \textit{momentum space} and to the fact that the Fermi surfaces for ``up'' and ``down'' spins are not well defined. The system still has a cylindrical axis of symmetry along the magnetization axis. We also show that the self energy has a universal structure which we determine based on the symmetries of the dipolar interaction and we explicitly calculated it in the Hartree-Fock approximation. We show that the bare magnetic moment of particles is renormalized due to particle-particle interactions and we give order of magnitude estimates of this renormalization effect. We estimate that the above mentioned dipolar effects are small but we discuss possible scenarios where this physics may be realized in future experiments.
 \end{abstract}
\pacs{03.75.Ss,05.30.Fk,75.80.+q,71.10.Ay}
\maketitle 

\section{Introduction}

 Interest in ultracold atoms with dipole-dipole 
interactions arise due to the long range and anisotropic nature of dipolar forces. 
If the dipolar Fermi gas is fully polarized the anisotropic form of the interactions leads to a distortion of the Fermi surface 
(FS) (Ref. \onlinecite{Miyakawa2008}, \onlinecite{Fregoso2009a} and Appendix A)
 and to a mixing of the collective excitations due to the structure of the Fermi liquid parameters.\cite{Fregoso2009a}
Although this system can be described in terms of the standard Landau theory of the Fermi liquid\cite{Baym1991} important changes are needed to 
account for the effects of the dipolar forces.\cite{Fregoso2009a,Ronen,Chan,Lin}
Recent theoretical work has revealed that gases of dipolar Fermi atomic systems can exhibit interesting and unconventional properties, 
including  novel ordered quantum liquid crystal states, such as biaxial nematic and ferronematic phases\cite{Fregoso2009a,Fregoso2009}.
Quantum liquid crystal phases have been conjectured to play an important 
role in the physics of strongly correlated systems\cite{Kivelson1998}. Recent experimental progress in trapping strongly 
magnetic dipolar Fermi gases in magnetic traps opens the way to study these interesting physical systems in the laboratory.\cite{Lu-2009}

Dipolar Fermi systems where 
spin degrees of freedom are allowed to play a dynamical role \cite{Fregoso2009,Shi,Wu} 
are much less studied. Magnetic dipole-dipole interactions, 
unlike electric dipole-dipole interactions, 
conserve total angular momentum,  spin + orbital ($\v{J}=\v{L}+\v{S}$),\cite{Pethick2008}, 
and hence mix the spin and orbital degrees of freedom. Interactions in 
dense nuclear matter with non-central forces are similar. In fact, 
many of the above effects have been studied theoretically in the nuclear 
physics literature.\cite{Olsson2002} 

In this work we consider, in perturbation theory, the effects of \textit{magnetic}
dipolar forces on a spin-\nicefrac{1}{2} Fermi system with a population imbalance 
of particles originally in eigenstates of the $\sigma^{z}$ Pauli matrix, 
corresponding to a partially polarized ferronematic state.\cite{Fregoso2009}
To first order in the interaction, he local spin quantization axis is not along $\uv{z}$, leading to an effective spin-orbit type interaction for the quasiparticles, whose
 self-energy develops a  spin structure in \textit{momentum space}.  In an earlier publication we showed that the qualitative physics of the ferronematic phase can be captured by the structure of the Fermi surfaces. The results we present here show that the nature of the ordered state at partial polarization has a complex, unconventional (i.e., a spin triplet particle-hole condensate anisotropic in orbital space), magnetic order associated with the dynamically generated spin-orbit type interactions.
Such a system can be simulated 
experimentally with two hyperfine levels of $^{163}$Dy in an optical trap at zero external magnetic field.\cite{Lu-2009}

We define the dipolar interaction parameter as $\lambda = \bar{n} \mu^2/\bar{\epsilon}_F$, where $\mu$ is the magnetic moment of the particles, 
$\bar{n}\equiv \bar{k}^{3}_F/(3\pi^2)$ and $\bar{k}_F$ is defined by the average chemical potential  
$\bar{\mu}\equiv(\mu_1+\mu_2)/2\equiv\bar{\epsilon}_F\equiv\hbar^2\bar{k}^{2}_F/2m$. Our calculations are to first order in $\lambda$, but the physical principles we 
describe are quite general. We do not consider here important questions such as pairing effects, collapse instabilities\cite{Fregoso2006}, temperature or optical trap 
effects. The physics we describe does not rely on trap parameters; such effects could be taken into account using Thomas-Fermi functionals.\cite{Ring1980}  Collapse 
instabilities, that we will not consider here,  require a finite dipolar coupling strength even for long range forces\cite{Fregoso2009} whereas the phenomena we study 
occurs for infinitesimal dipolar couplings. 

Before presenting the details we summarize the physical picture 
that emerges as follows. A fixed difference of particles in the eigenstates of $\sigma^{z}$ creates a net magnetization of the system which then
acts, as far as static processes are concerned, as a uniform external magnetic field along $\uv{z}$. In the absence of magnetic field the dipole interaction conserves 
the spin + orbital angular momentum ($\v{S}+\v{L}$). Crucially, in the presence of a net magnetization the colliding particles still conserve the $z$-projection of the 
total angular momentum, $J_z = L_z + S_z$ along the magnetization. Recall that the magnetic dipole interaction is an $L=2$ object under rotations of the spatial 
coordinates\cite{Pethick2008} and that, unlike contact, Coulomb or exchange interactions, the dipolar forces can flip the spin of one of the two interacting particles. To 
conserve angular momentum the missing momentum is taken from the orbital degrees of freedom. In a many-body system this means that any given particle will ``feel'' 
an average magnetic field due
to the interaction with other particles. In other words, the mean field is a vector not a scalar. 
This molecular magnetic field is capable of flipping the spin of the spin-\nicefrac{1}{2} and hence it must point away from the 
the $z$-axis. Moreover since $J_z$ must be conserved the change in spin angular momentum due to a spin flip process 
must be accompanied by a corresponding change in orbital (momentum space) angular momentum. In this way we arrive 
at the conclusion that the distribution in momentum space of the particles must be anisotropic and that 
the spin state of the particle will depend on its velocity!

Since the local spin quantization axis is not parallel to the polarization direction $\uv{z}$, the $\v{k}$-dependent spin state 
of the particles becomes a linear combinations of eigenstates of $\sigma^{z}$, i.e., the spins tilt. 
Interestingly, we find that neither the tilting angle nor the momentum distribution of the quasiparticles 
are isotropic in momentum space, and vary with direction 
as $\sim |Y_{2\pm 1}(\uv{k})|$ and $Y_{20}(\uv{k})$ respectively because of the 
mixing  of orbital and spin degrees of freedom.
This can be described in perturbation 
theory as arising from two processes: a) $\left|\up \right\rangle \to \left|\up \right\rangle$ (no spin flip), with $\Delta L_z=0$; 
since we consider particles with no internal structure, the quasiparticle self-energy acquires a correction
of the form $Y_{20}(\uv{k})$, b) $\left|\up \right\rangle \to \left|\down \right\rangle$
(spin flip) with $\Delta L_z=\mp 1$, the quasiparticle self-energy gets a correction 
of the form $|Y_{2,\pm1}(\uv{k})|$. Therefore the self-energy now has an off-diagonal momentum-dependent component. These changes in the structure of the self-energy lead to qualitative modifications to the quasiparticle distribution functions and of the Fermi surface. In particular it is no longer possible to define separate Fermi surfaces for up and down fermions independently.

This work is organized as follows. In section \ref{sec:single_particle_states} we investigate the structure of the single particles states and show how the orbital and spin degrees of freedom become entangled due to the presence of non-diagonal elements in the self energy of particles.  
In section \ref{sec:renormalization_mag_moment} we show that particle interactions lead to a renormalization of the bare magnetic moment. 
In section \ref{sec:one_body_density_matrix} we compute the occupation number in momentum space noting that this quantity is a $2\times 2$ tensor in spin space. In section \ref{sec:discussion} we provide possible scenarios where dipolar effects may be observed. Finally the appendices provide the details of the self energy computations for the fully polarized and partially polarized case (Appendix \ref{app:self_energy_fully_polarized} and \ref{app:self-energyS11andS12}), an intuitive explanation of the FS distortion (Appendix \ref{app:distortion}) and details of the Fourier transform of the dipolar interaction (Appendix \ref{app:fourier}).

\section{Structure of the quasiparticle self-energy}
\label{sec:single_particle_states}

 The excitations of a gas of fermions at finite density are quasiparticles with the same quantum numbers as the fermions. However their dynamics is strongly affected by their mutual interactions. These physical effects are captured by the quasiparticle self-energy.
In the presence of dipole-dipole 
interactions the form of the self energy is determined by
symmetry arguments. We thus require that the quasiparticle self energy be invariant 
under simultaneous rotations of the net magnetic field and the quasiparticle momentum, which leads to an expression of the form
\beq
\Sigma^{ij}(\v{k},\omega) = \Sigma_0(\mathbf{k},\omega) \delta_{ij} + \frac{\Sigma_T(\mathbf{k},\omega)}{2}(3\uv{k}_{i}\uv{k}_{j} -\delta_{ij}) 
\label{enq:self-energy-general-expression}
\eeq
where $\Sigma_0$ and $\Sigma_T$ are functions of the wave vector and frequency, $i=x,y,z$ is the direction of the effective magnetic field and $j=x,y,z$ 
is the direction of the spin polarization. The form of this expression is valid to all orders in perturbation theory.  

Without loss of generality, we consider a microscopic Hamiltonian with a population 
imbalance of particles in eigenstates of $\sigma^{z}$, {\it i.e.\/} we will assume a state with a fixed total polarization. The Hamiltonian is 
\begin{eqnarray}
\hat{H}&=&\sum_{\mathbf{k},\alpha} \epsilon_{\mathbf{k}} c^\dagger_{\mathbf{k}.\alpha} c_{\mathbf{k},\alpha}+g\sum_{\mathbf{k}}n_{\mathbf{k},\up}n_{\mathbf{k}\down}
\nonumber\\
+&& \frac{1}{2V}\sum_{\mathbf{q}} \hat{S}^{i}(\v{q})V_{ij}(\mathbf{q}) \hat{S}^{j}(-\mathbf{q}) 
\label{eqn:dipole-dipole-hamiltonian}
\end{eqnarray}
where $\epsilon_{\textbf{k}}=\frac{\hbar^2 \textbf{k}^2}{2m}$ are the singe particle energies, $n_{\mathbf{k},\up}$ and $n_{\mathbf{k},\down}$ are the occupation numbers of the single particle states with momentum $\mathbf{k}$ with both spin projections, and $g$ is the strength of the repulsive short-range s-wave interaction. The last term is the dipolar interaction. In Fourier space  it is given by (see Appendix \ref{app:fourier})
\begin{equation}
V_{ij}(\v{q}) = (4\pi\mu^2/3)(3\uv{q}_i\uv{q}_j - \delta_{ij})
\label{eqn:FT_dipole}
\end{equation}
where 
$\uv{q}$ is the unit vector in the direction of $\v{q}$, $\mu \equiv\gamma \hbar/2$ is the 
bare magnetic moment of particles and $\gamma$ the gyromagnetic ratio, e.g., 
for electron spins, $\gamma = 2\mu_B$. Repeated indices are to be summed over.
$\hat{S}^{i}(\v{q})=\sum_{\v{k}} c^{\dagger}_{\v{k}+\v{q},\alpha} \sigma^{i}_{\alpha\beta} c_{\v{k},\beta}$ 
is the spin density in momentum space. $\sigma^{i}$ are Pauli matrices and 
$\alpha$, $\beta$ are spin indexes that label the eigenstates of $\sigma^{z}$. 

In what follows we will assume that either the system is prepared in a state of finite polarization or that $g$ is large enough for the ground state to be a ferronematic with finite polarization (as discussed in Ref.[\onlinecite{Fregoso2009}]), and compute the fermion self-energy perturbatively in the dipolar interaction.
To first order, the self energy is given by (see Fig.1)
\beq
\hbar \Sigma_{\alpha\beta}(\v{k}) &=& -\int \frac{d^3 k'}{(2\pi)^3} \frac{1}{\hbar \beta}\sum_{n'} \mathrm{e}^{i \omega_{n'} 0^{+}} \nn \\ 
&&\times\; \G^{0}_{\delta\gamma}(\v{k}',i\omega_{n'}) V_{ij}(\v{k}-\v{k}') \sigma^{i}_{\alpha\delta}\sigma^{j}_{\gamma\beta} 
\label{eqn:self-energy-matrix-form}
\eeq
Explicitly (see appendix \ref{app:self-energyS11andS12}) 
\beq
\hbar\Sigma_{11} &=& -\hbar\Sigma_{22} = -f(k) (3 \uv{k}^2_z -1 ) \nn \\
\hbar\Sigma_{12} &=& \hbar\Sigma_{21}^{*} = -f(k) 3\uv{k}_z ( \uv{k}_x + i \uv{k}_y).
\label{eqn:self_energy_matrix_elements}    
\eeq 
Being a $2\times 2$ matrix we can expand $\hbar \Sigma_{\alpha\beta}(\v{k})$ in Pauli matrices, 
$\hbar\Sigma_{\alpha\beta}= \Sigma^{zi}(\v{k})\sigma^{i}_{\alpha\beta}$, where we found 
$\Sigma^{zi}(\v{k}) = -f(k)(3 \uv{k}_{i}\uv{k}_{z} -\delta_{i,z})$. 
The angles of $\v{k}$ are defined in Fig. \ref{fig:local-quatization-axis}. Comparing with the general expression Eqn. \eqref{eqn:self-energy-matrix-form}, we conclude that in the Hartree-Fock approximation
\beq
\Sigma_0 &=&0 \nn \\
\Sigma_{T} &=& -2 f(k).  
\label{eq:sigmas}
\eeq
\begin{figure}[hbt]
\begin{center}
\subfigure{\includegraphics[width=0.5\textwidth]{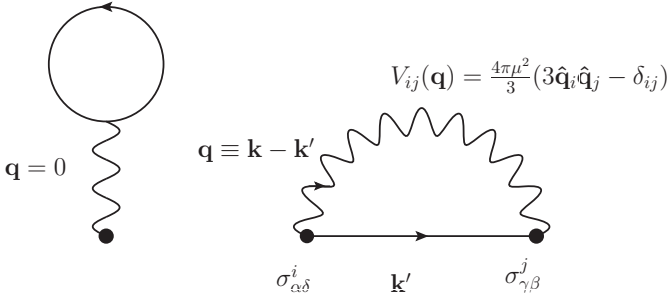}}
\caption{Feynman diagrams for Hartree (left) and Fock (right) terms of the self energy}
\label{fig:HF_feynman_diagram}
\end{center}
\end{figure} 

\begin{figure}[hbt]
\begin{center}
\subfigure{\includegraphics[width=0.45\textwidth]{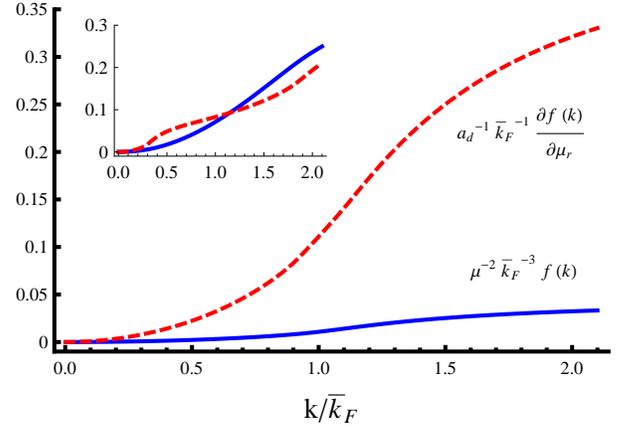}}
\caption{(color online) characteristic form of $f(k)$(blue-solid) and its derivative (red-dashed) for two values of $\mu_r$ and $\bar{\mu}$,
Eqn. \eqref{eqn:self-energy-magnitude}. Main panel: $k_{F1}=1.1$, $k_{F2}=0.9$, inset: $k_{F1}=1.8$, $k_{F2}=0.3$ in units of $\bar{k}_F$.}
\label{fig:HF_feynman_diagram_and_fk}
\end{center}
\end{figure} 

We assumed a fixed population imbalance of particles in the two 
eigenstates of $\sigma^{z}$, by introducing two chemical potentials $\mu_1$ and $\mu_2$.  
$G^0_{\alpha\beta}({\v{k},i\omega_{n}}) = \delta_{\alpha\beta}/(i\omega_{n} - \xi_{\v{k}\alpha})$ 
is the free temperature Green function, where 
$\hbar \xi_{\v{k}\alpha} = \epsilon^0_{\v{k}} - \mu_\alpha$ is the bare single particle dispersion 
relation and $\epsilon^{0}_{\v{k}} = \hbar^2 k^2/2m $. In Eq.\eqref{eq:sigmas}
$f(k)$ is a smooth monotonic function of the magnitude of the wave vector: 
\beq
f(k)&=& \frac{2\mu^2}{\pi} \int^{\infty}_0 {k'}^{2} dk' \int^{\infty}_{0} \frac{dr}{r} j_{2}(kr) j_{0}(k'r) \nn \\
&&\times \left[ n_{F}(\epsilon^0_\v{k'} - \mu_1) -n_{F}(\epsilon^0_\v{k'} -\mu_2) \right]
\label{eqn:self-energy-magnitude}
\eeq
where $n_F(x)=[\exp{(\beta x)}+1]^{-1}$ is the Fermi
function. The self energy vanishes for an equal number of particles in each 
eigenstate of $\sigma^{z}$. This means that dipole torques exist only when there is a 
net magnetization in the system, as we would expect.

The diagonal matrix elements of the self energy, proportional to $\pm(3\cos^2\theta_{\v{k}}-1)$, give anisotropic renormalization to the single particle energies. It was show that for partial polarization the majority-component FS is elongated while the minority-component FS compression along the polarization axis.\cite{Fregoso2009} The novel feature of Eqn. \ref{eqn:self_energy_matrix_elements} is its non-diagonal matrix elements. They are not zero because the dipole interaction can flip \textit{one} of the two interacting particles. This means that the mean field acting on a given particle's spin can flip it. In turn this imply that such an effective molecular vector field is not along $z$-axis. The structure of the terms: $\Sigma_{12}(\v{k})\sim Y_{2,-1}(\uv{k})$, and $\Sigma_{21}(\v{k})\sim Y_{2,1}(\uv{k})$ of the non-diagonal matrix elements can be understood as a consequence of angular momentum conservation. Since the magnetic dipole-dipole interaction conserves $J_z=L_z+S_z$ in the presence of a net polarization along $z$-axis, a change in the spin angular momentum $\Delta S_z=\pm 1$ in accompanied by a corresponding change in the orbital angular momentum $\Delta L_z=\mp 1$. 

The particle energy, $(\epsilon_{\v{k}})_{\alpha\beta}$, have a \textit{tensorial}
structure and is given by the equation $\hbar \xi_{\v{k}\alpha} \delta_{\alpha\beta} + \hbar\Sigma_{\alpha\beta}(\v{k}) \equiv (\epsilon_{\v{k}})_{\alpha\beta} - \bar{\mu}\delta_{\alpha\beta}$. Expanding in Pauli matrices we obtain 
\beq
(\epsilon_\v{k})_{\alpha\beta} \equiv \epsilon^{0}_{\v{k}}\delta_{\alpha\beta} +
\epsilon^{i}_{\v{k}} \c \sigma^{i}_{\alpha\beta},
\label{eqn:def-local-field}
\eeq
we see that there is an the effective magnetic field, $\v{h}_{\v{k}}$, given by $\epsilon^{i}_{\v{k}} = \mu h^{i}_{\v{k}}$, that acts on the particle's magnetic moment. This  $\v{k}$-dependent field has a contribution from the  external field long $\uv{z}$ (i.e.,  $\mu_r$) and a molecular contribution due to interactions with other particles. This field is necessarily \textit{not} along $\uv{z}$. We explicitly write the components $\epsilon^{i}_{\v{k}}$ in Fig. \ref{fig:local-quatization-axis} and we plotted $\v{h}_{\v{k}}$ in Fig. \ref{fig:effective-field}. 
\begin{figure}[hbt]
\subfigure{\includegraphics[width=0.35\textwidth]{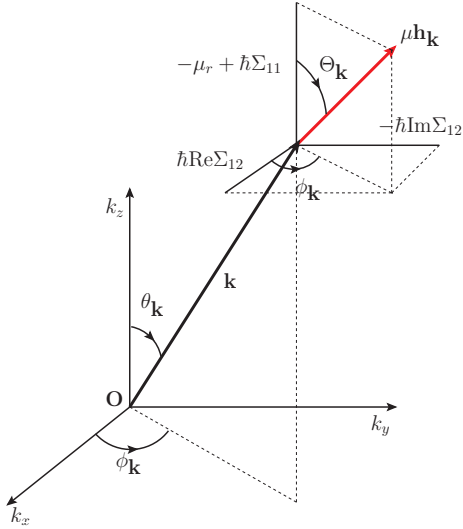}}
\caption{(color online) Effective magnetic field, (red arrow), at point $\v{k}$ in momentum space. 
The local spin quantization axis is tilted with respect to $\uv{z}$ because the dipole-dipole interaction induce
non-diagonal matrix elements, $\Sigma_{12}(\v{k})$, in the self energy, Eqn. \eqref{eqn:self_energy_matrix_elements}.}
\label{fig:local-quatization-axis}
\end{figure} 

\section{Quasiparticle magnetic moment}
 \label{sec:renormalization_mag_moment}

\begin{figure}
\subfigure{\includegraphics[width=0.30\textwidth]{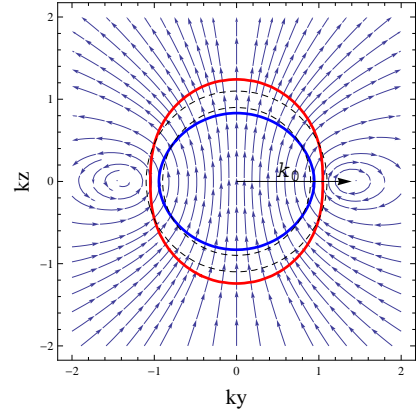}}
\caption{(color online) 2D view of points in $\v{k}$-space where $E_{\v{k}\pm}=\bar{\mu}$(red and blue). 
FS's of free system: $\epsilon^{0}_{\v{k}}=\bar{\mu} \pm \mu_r$(dashed lines). 
Superposed are stream lines of magnetic moment field carried by particles 
in momentum space, Eqn. \ref{eqn:mag-moment-general-form}. This field resembles 
the magnetic field produced by an electric current circulating counterclockwise 
in a ring placed in the horizontal plane. Values for these plots are $k_{F1}=1.1$ 
and $k_{F2}=0.9$, in units of $\bar{k}_{F}$, and   
$\lambda =0.3<< 1$.}
\label{fig:effective-field}
\end{figure}

As we remarked above the problem of a system with a fixed population imbalance is equivalent to have a static magnetic field in the system. Hence it is of interest to know what is the magnetic response of the system. In a small external field we define the effective magnetic moment of the particle as
\beq
\tilde{\mu}_{ij}(\v{k})\equiv -\frac{\partial \epsilon^{i}_{\v{k}}}{\partial h^{j}_{ext}} = 
-\mu\frac{\partial \epsilon^{i}_{\v{k}}}{\partial \mu^{j}_r}.
\label{eqn:def-magnetic-moment}
\eeq
Where we defined the external field along $\uv{z}$ by $\mu h^{z}_{ext} = \mu_r$. The effective magnetic moment of the particles is not the bare magnetic moment $\mu$. In fact, when dipole-dipole interactions are present we can guess the form of the magnetic moment carried by the particle by symmetry arguments.\cite{Olsson2002}. We require that the magnetic moment be invariant under simultaneous rotations of the magnetic field and the momentum of the particle, i.e., 

\beq
\tilde{\mu}_{ij}(\v{k},\omega) = \tilde{\mu}_{0} \delta_{ij} + \frac{\tilde{\mu}_T(\v{k},\omega)}{2}(3 \uv{k}_i\uv{k}_j -\delta_{ij} )
\label{eqn:mag-moment-general-form}
\eeq
where we find $\tilde{\mu}_0 =\mu$ and $\tilde{\mu}_T=2 \mu (\partial f(k)/\partial \mu_r)$.
$f(k)$ is given by Eqn. \ref{eqn:self-energy-magnitude}.
For the values used in Fig. \ref{fig:effective-field} we find $\tilde{\mu}_T/\tilde{\mu}_0 \sim 0.3$ 
at the outer FS. For small dipolar couplings, where our microscopic calculations are accurate, 
the occupied states in momentum space lie outside the singularity, see Fig. \ref{fig:effective-field}.
The line singularity is a ring in the $k_x$,$k_y$-plane with radius, $k_0$, given by 
$\mu_r = f(k_0)$. This wave vector, $k_0 \sim (\mu_r/\mu^2)^{1/3}$,  
is perpendicular to the polarization axis, assumed long $\uv{z}$, 
and naturally introduces a length scale, $1/k_0$ into the system. In the regime were our calculation is valid the length scale
is much sorter than the inter-particle distance, $1/k_0 < 1/\textrm{max}(k_{F1},k_{F2})$ and
hence it is related to microscopic effects not considered by our long-distance microscopic dipole-dipole 
Hamiltonian. For strong dipolar couplings, our calculation suggests that the radius 
of the ring penetrates the FS's. In this case, particles would occupy spin states with non-trivial 
topological structure. At a critical coupling the system may eventually acquire a spatial modulation.\cite{comment1}. 
The strongly coupled regime, $\lambda \rightarrow \infty$, can be accessed 
by increasing the dipole interaction or by reducing the kinetic energy. The latter possibility can be experimentally 
engineer in the laboratory by placing the atoms in an optical lattice.

\section{one-body density matrix}
\label{sec:one_body_density_matrix}
In this section we compute the one-body density matrix in Hartree-Fock approximation. The distribution function of particles,  $(n_{\v{k}})_{\alpha\beta}$, is now a $2\times 2$ matrix in spin space. Dyson equation, $(\G^{-1})_{\alpha\beta} = (i\omega_n -\xi_{\v{k}\alpha})\delta_{\alpha\beta}-\Sigma_{\alpha\beta}(\v{k},\omega)$, with Hartree-Fock self energy corrections provide a conserving approximation\cite{Kadanoff1989} to the Green function. Defining $\hbar \xi_{\v{k}\pm} = E_{\v{k}\pm} - \bar{\mu}$, an explicit calculation gives 
\beq
\G_{11}(\v{k},i\omega_n) &=&  \frac{u^{2}_{\v{k}}}{i\omega_n- \xi_{\v{k}+}} + \frac{v^{2}_{\v{k}}}{i\omega_n- \xi_{\v{k}-}} \nn \\ 
\G_{22}(\v{k},i\omega_n) &=& \frac{v^{2}_{\v{k}}}{i\omega_n- \xi_{\v{k}+}} + \frac{u^{2}_{\v{k}}}{i\omega_n- \xi_{\v{k}-}} \nn \\
\G_{12}(\v{k},i\omega_n) &=& u_{\v{k}} v_{\v{k}} e^{-i\phi_{\v{k}}}\left[\frac{1}{i\omega_n- \xi_{\v{k}+}} - \frac{1}{i\omega_n- \xi_{\v{k}-}}\right] \nn \\
\G_{21}(\v{k},i\omega_n) &=& u_{\v{k}} v_{\v{k}} e^{i\phi_{\v{k}}}\left[\frac{1}{i\omega_n- \xi_{\v{k}+}} - \frac{1}{i\omega_n- \xi_{\v{k}-}}\right] \nn \\
&& 
\label{eqn:green_functions}
\eeq
where $E_{\v{k}\pm} = \epsilon^{0}_\v{k} \pm \big[(\hbar\Sigma_{11}(\v{k})-\mu_r)^2 + |\hbar\Sigma_{12}(\v{k})|^2\big]^{1/2}$ and $u_{\v{k}} = \cos(\Theta_{\v{k}}/2)$, $v_{\v{k}} = \sin (\Theta_{\v{k}}/2)$ and $0<\Theta_{\v{k}}\leq \pi$ is defined in Fig. \ref{fig:local-quatization-axis}. $u^{2}_{\v{k}}$ is the probability of up-spin propagation and  $v^{2}_{\v{k}}$ is the probability of down-spin propagation. Equating $\G= \G_0 \mathbf{1} + \G_1 \sigma^x+ \G_2 \sigma^y+ \G_3 \sigma^z$ and collecting like terms we can write in compact form
\beq
\G(\v{k},i\omega_n) = \frac{(i\omega_n - \xi_{\v{k}M})\mathbf{1} + \xi_{\v{k}r}\uv{h}_{\v{k}}\cdot{\boldsymbol \sigma}}{(i\omega_n - \xi_{\v{k}+})(i\omega_n - \xi_{\v{k}-})}
\eeq
where $\xi_{\v{k}M} \equiv (\xi_{\v{k}+} +\xi_{\v{k}-})/2 = (\epsilon^{0}_{\v{k}} -\bar{\mu})/\hbar$ and $\xi_{\v{k}r} \equiv (\xi_{\v{k}+} -\xi_{\v{k}-})/2$. $\uv{h}_{\v{k}}$ is a unit vector along the effective magnetic field parametrized by the polar angles $\Theta_\v{k},\phi_\v{k}$, see Fig. \ref{fig:local-quatization-axis}. $\mathbf{1}$ is the $2\times 2$ identity matrix. In our case the one-body density matrix is given by $(n_{\v{k}})_{\alpha\beta} \equiv \int (d\omega/2\pi) n_{F}(\omega) \rho_{\alpha\beta}(\v{k},\omega)$, where $\rho_{\alpha\beta}$ is a $2 \times 2$ tensor of spectral functions and $n_{F}(\omega) = [\exp{(\beta\omega)} + 1]^{-1}$. The spectral functions are given by 
\beq
i \rho_{\alpha,\beta}(\v{k},x) &=& \G(\v{k},\omega_n)\big|_{i\omega_n = x-i0^{+}} \nn \\
&&\hspace{30pt}- \G(\v{k},\omega_n)\big|_{i\omega_n = x+i0^{+}} 
\eeq
In Hartree-Fock, we obtain two simple delta functions corresponding to the poles of 
the Green functions. It is useful to expand the  density matrix in Pauli matrices, 
\beq
(n_{\v{k}})_{\alpha\beta} = \bar{n}_{\v{k}}\delta_{\alpha\beta} +  
\frac{1}{\hbar}\left\langle S^{i}_{\v{k}}\right\rangle \c \sigma^{i}_{\alpha\beta}.
\label{eqn:Pauli-expandsion-density-matrix}
\eeq
where  $\bar{n}_{\v{k}}\equiv (1/2)\textrm{tr}[(n_{\v{k}})]$ is the 
average occupation of state with momentum $\v{k}$, and
$\left\langle S^{i}_{\v{k}}\right\rangle \equiv \hbar \c n^{i}_{\v{k}}\equiv
(\hbar/2)\textrm{tr}[\sigma^{i}(n_{\v{k}})]$  is the 
$i$-th spin density component. $\left\langle S^{i}_{\v{k}}\right\rangle$ 
has a clear physical interpretation namely, it is a vector field with polar 
angles defined in Fig. \ref{fig:local-quatization-axis} and plotted in 
Fig. \ref{fig:effective-field}, which gives the $\v{k}$-dependent quantum 
mechanical average of the spin state of the particle. Explicitly,
\beq
\bar{n}_{\v{k}} &=& \frac{1}{2}\left[ n_{F}(E_{\v{k}+}) + n_{F}(E_{\v{k}-})\right] \nn \\
\left\langle S^{x}_{\v{k}}\right\rangle &=& \hbar\c u_{\v{k}} v_{\v{k}}\cos\phi_{\v{k}}\c 
\left[ n_{F}(E_{\v{k}+} ) - n_{F}(E_{\v{k}-} ) \right] \nn \\
\left\langle S^{y}_{\v{k}}\right\rangle &=& \hbar\c u_{\v{k}} v_{\v{k}}\sin\phi_{\v{k}}\c
\left[ n_{F}(E_{\v{k}+} ) - n_{F}(E_{\v{k}-} ) \right]\nn \\
\left\langle S^{z}_{\v{k}}\right\rangle&=& \frac{\hbar}{2}(u^{2}_{\v{k}} - v^{2}_{\v{k}}) 
\left[ n_{F}(E_{\v{k}+} ) - n_{F}(E_{\v{k}-} ) \right].  \nn 
\eeq
Note that in these expressions we define $n_F(x) = [\exp{(\beta x - \bar{\mu})}+1]^{-1}$. 
Since $n^{x}_{\v{k}}\neq 0$ and $n^{y}_{\v{k}}\neq 0$, there is a 
non-zero average probability that the spin is pointing along the $xy$-plane 
in the region between the blue and red curves in Fig. \ref{fig:effective-field}.
One can check that 
\beq
\sum_{\v{k}}\left\langle S^{x}_{\v{k}}\right\rangle=\sum_{\v{k}}\left\langle S^{y}_{\v{k}}\right\rangle=0.
\eeq
i.e., there is no overall spin magnetization of the gas in the $xy$-plane. 
The system has cylindrical symmetry along the magnetization axis,  see Fig. \ref{fig:effective-field}. 

The net spin magnetization is along $\uv{z}$ and is given by 
$\sum_{\v{k}}\left\langle  S^{z}_{\v{k}}\right\rangle$.
The average occupation number of state $\v{k}$ and spin projection 
along $\uv{z}$:  $n_{\v{k}\up}\equiv\bar{n}_{\v{k}} + n^{z}_{\v{k}}$ and 
$n_{\v{k}\down}\equiv\bar{n}_{\v{k}} - n^{z}_{\v{k}}$, are
\begin{figure}[hbt]
\subfigure{\includegraphics[width=0.35\textwidth]{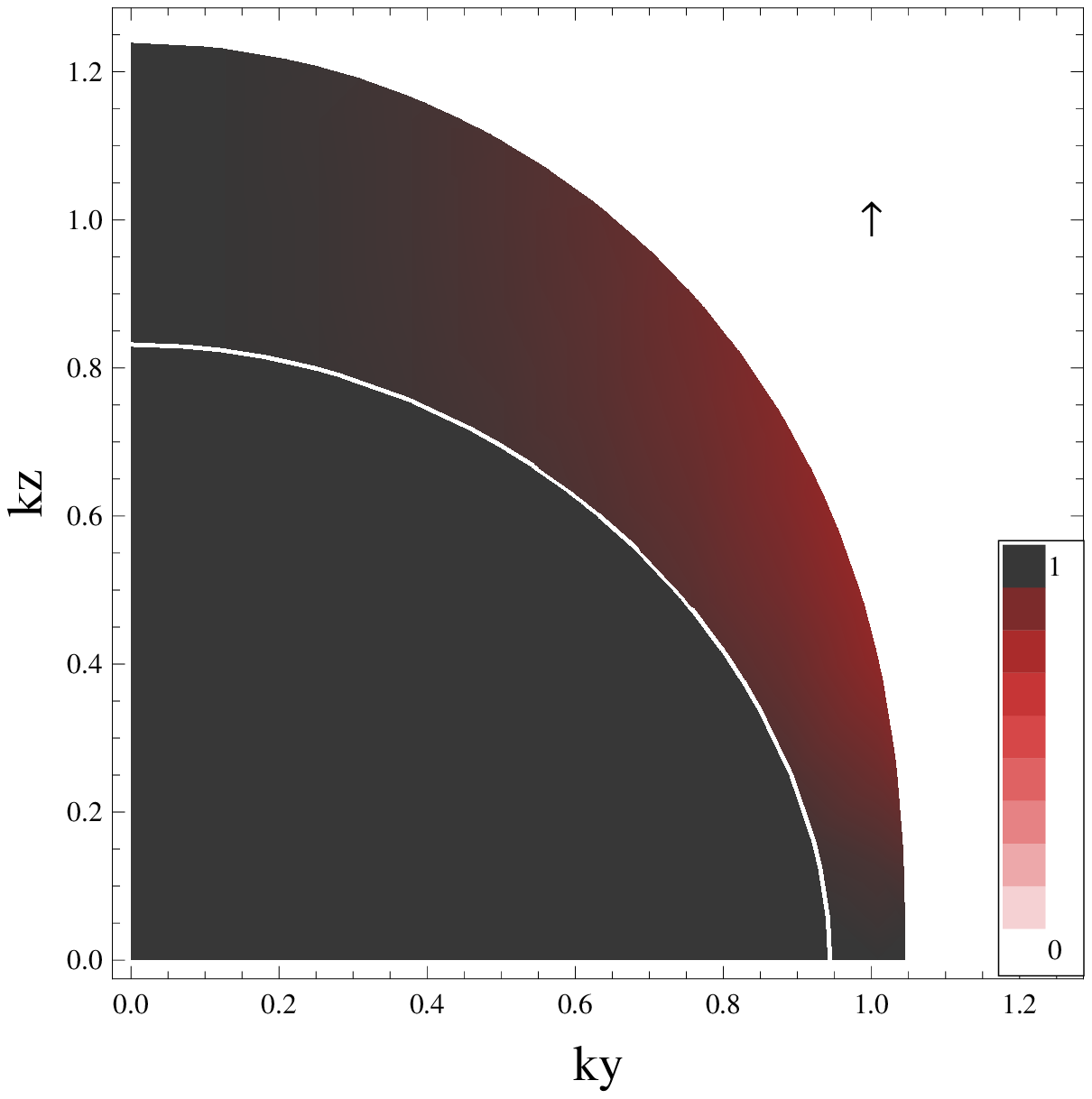}}
\subfigure{\includegraphics[width=0.35\textwidth]{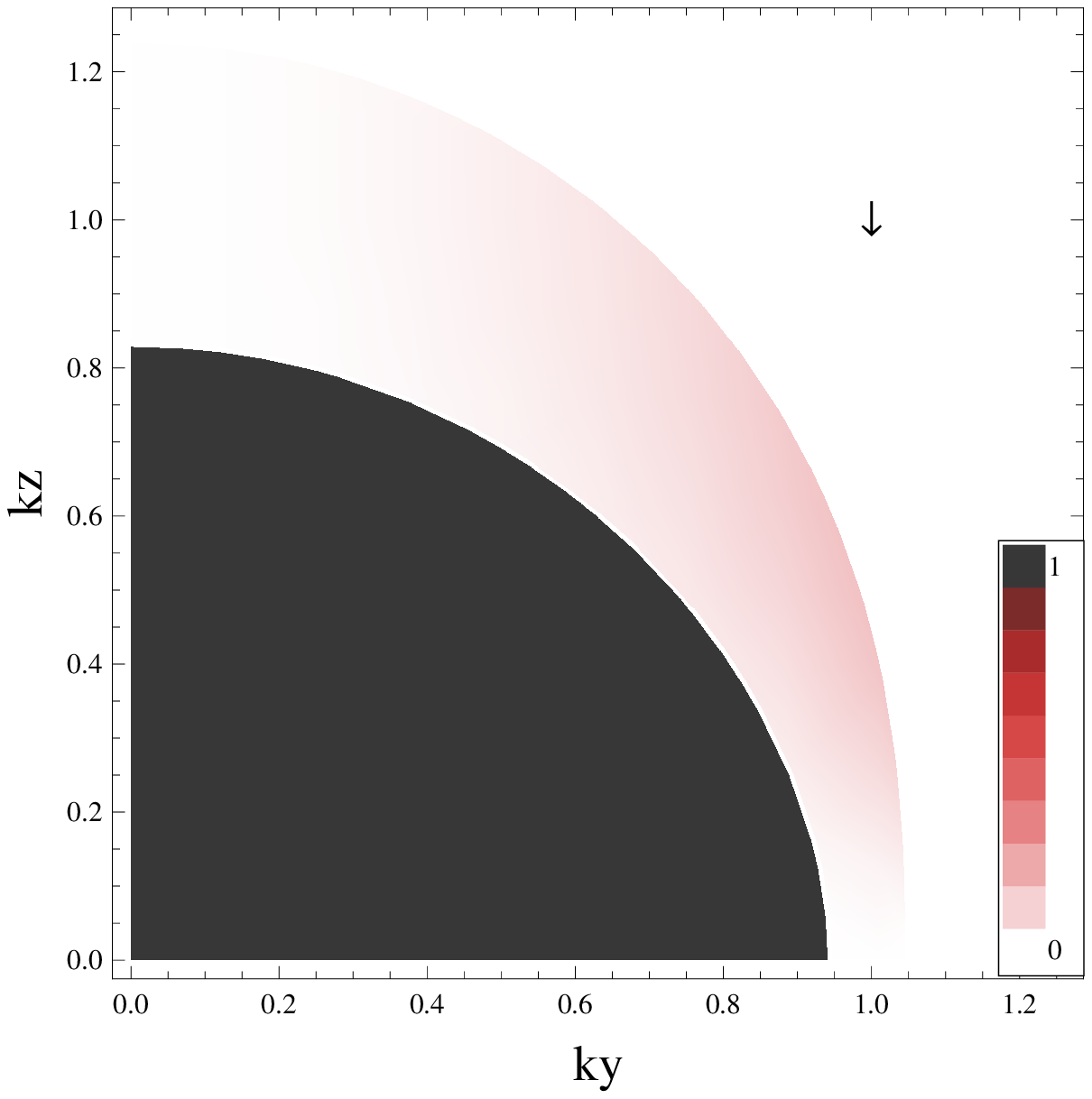}}
\caption{(color online) density plot of the occupation probability in momentum space of $z$-projection of spin, $n_{\v{k}\up}$ (a) and 
$n_{\v{k}\down}$ (b), (Eqn. \eqref{eqn:momdistribution-up-and-down}). Values for these plots are the same as those of Fig. \ref{fig:effective-field}.}
\label{fig:mondistribution-densityplot}
\end{figure} 
\begin{figure}[hbt]
\subfigure{\includegraphics[width=0.35\textwidth]{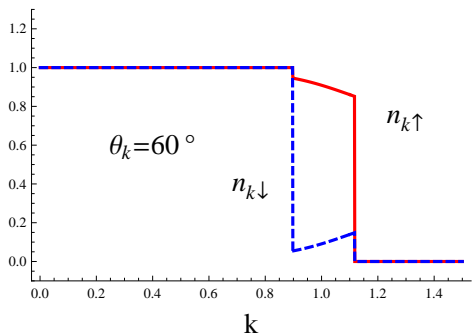}}
\caption{(color online) momentum distribution along a line with polar angle 
$\theta_{\v{k}} = 60\deg$  (see Fig.\ref{fig:local-quatization-axis}) for $n_{\v{k}\up}$(red-solid) $n_{\v{k}\down}$(blue-dashed).}
\label{fig:cut_mom_distribution}
\end{figure} 
\beq
n_{\v{k},\up} = u^{2}_{\v{k}} n_{F}(E_{\v{k}+}) + v^{2}_{\v{k}} n_{F}(E_{\v{k}-}) \nn  \\
n_{\v{k},\down} = v^{2}_{\v{k}} n_{F}(E_{\v{k}+}) + u^{2}_{\v{k}} n_{F}(E_{\v{k}-}).
\label{eqn:momdistribution-up-and-down} 
\eeq
Fig. \ref{fig:mondistribution-densityplot} is a density plot of the occupation probabilities of $z$-component of the spin in momentum space. Note that the average occupation of spin states is not uniform in momentum space as it would in a non-interacting Fermi gas in which case there would be two concentric spheres uniformly occupied with the ``up'' or ``down'' eigenstates of $\sigma^{z}$. Note also that dipole-dipole interactions cause the spins to tilt predominantly along the diagonals, see also Fig. \ref{fig:effective-field}, in such a way that measurement of quantum spin state will yield ``down'' with non-vanishing probability even in regions in momentum space originally occupied only by spins pointing up. 

Eqn. \eqref{eqn:momdistribution-up-and-down} implies that the ground state wave function which evolves adiabatically from the non-interacting system is of the form 
\beq
|\Psi~\rangle_{D} = \prod_{\v{k} \in \Omega_{\v{k}_{F}}} \left(u_{\v{k}} c^{\dagger}_{\v{k},\up}
+  v_{\v{k}} c^{\dagger}_{\v{k},\down} \right)|0 \rangle \nn
\eeq
where $u_\v{k}$ is the $\v{k}$-dependent amplitude for the spin  to be the eigenstate ``up'' of $\sigma^z$ and $v_\v{k}$ the $\v{k}$-dependent amplitude for the spin to be the eigenstate ``down''. In this work we computed these coefficients to first order in the dipole interaction. As we see the spin states of particles are linear combinations of eigenstates of $\sigma^z$. The true FS, the one which separates the occupied from unoccupied single particle states, is given by the red curve in Fig. \ref{fig:effective-field} i.e., by the values of $\v{k}_F$ that satisfy $E_{\v{k}+} -\bar{\mu}\equiv \epsilon^{0}_{\v{k}_F} + \big[(\hbar\Sigma_{11}(\v{k})-\mu_r)^2 + |\hbar\Sigma_{12}(\v{k})|^2\big]^{1/2} - \bar{\mu}=0$. There are sharp discontinuities in momentum space, see Fig. \ref{fig:mondistribution-densityplot}. In this sense they define elliptically distorted FS's such as those found in Ref.[\onlinecite{Fregoso2009}]. We comment that as the polarization of the gas increases $v_{\v{k}}$ decreases and for a fully polarized gas we recover the elliptical shape, red curve in Fig. \ref{fig:effective-field} found in 
Refs. [\onlinecite{Fregoso2009a,Miyakawa2008,Sogo2009}].

\section{Discussion}
\label{sec:discussion} 

We have used perturbation theory to find the magnetic structure of the ground state of an itinerant, magnetized, spin-\nicefrac{1}{2} Fermi gas with dipole forces. Our calculations are the first step in the analogous perturbative calculations of Gell-Mann and Br\"uckner of the energy of electron gas and hence are of general theoretical interest in themselves. However, we have ultracold atomic gases with strong magnetic moments, such as $^{163}$Dy ($10\mu_B$) as a specific candidate\cite{Lu2010}. The dipole interaction of two magnetic dipoles in a Fermi gas with density $10^{13} cm^{-3}$ is $\sim 10$nK and may be too small compared with other energy scales in the system. We expect that it is an experimental challenge so observe directly dipole effects. In fact, direct evidence of this unconventional magnetism may become possible using a variation of NMR experiment on ultracold dipolar gases as in the superfluid phase just as in $^3$He-A phase\cite{Leggett1975}. We briefly consider the following scenarios of possible realization of dipolar effects in Fermi systems.

The ``spin'' states could be two hyperfine levels of $^{163}$Dy in an optical trap. In its ground state,$^{163}$Dy  has spin, orbital, nuclear and total electronic angular momentum of $S=2$, $L=6$, $I=5/2$ and $J=8$ respectively. The total angular momentum of the atom can be $F=11/2,...,21/2$. However, the energy gaps are several orders of magnitude larger that the typical dipole-dipole energy per particle attained with current experimental densities and its possible that the spin structures we describe here are washed out. However, since we are operating in an optical trap, it may be possible to use the low field limit from the \textit{same} hyperfine-Zeeman manifold, in which case, the splitting between hyperfine-Zeeman states is not large. Now, if we operate in the high field limit (as we may want $a_s=0$, the Feshbach resonance in Dy) then clearly the effects of dipole interaction are suppressed. At the time of writing little is known about the theory Feshbach resonances in Dy. In a closely related problem we assumed that the imbalanced Fermi gas was metastable. A possible stabilizing mechanism is provided by a Stoner-type of effective interaction\cite{Gyu-Boong2009,Dong} which was not considered in the present work. Another possibility is to choose two hyperfine states for which conservation of angular momentum would strongly limit the phase space available for particles to decay into. If this is true for Dy then there is no need to invoke the Stoner mechanism.

A second possibility is the use of an optical lattice loaded with hyperfine states of $^{163}$Dy. This method has certain advantages as far as stability is concerned \cite{Zhang} and might also make the dipolar effects much more pronounced. By working with a small filling fraction to avoid a large same-site interaction and umklapp processes it may be possible to make the dipole interaction be the dominant energy scale  

Another important experimental question is the possible presence of external stray magnetic fields. In the present work we already assumed a fixed external Zeeman field. Hence, the same considerations apply as regards to the observability of dipolar effects under stray magnetic fields. In particular if $E_{ex}< E_{d}$, the external field will pin the overall axis of symmetry of the system but will not modify significantly the spin structures described here. We only considered zero temperature effects. We known that temperature effects decrease the FS distortions\cite{Fregoso2010,Fregoso2009a,Endo,Kestner}. In general we expect the spin structures we found to be very sensitive to temperature. For example, for current attainable densities($10^{13} cm^{-3}$) the dipolar energy per particle is $E_{d}\sim 10nK$. Boundary effects are also known to be very important for long range forces such as dipolar. If a degenerate dipolar Fermi gas is realized one will obtain images that resemble Fig. \ref{fig:mondistribution-densityplot} in time-of-flight experiments. 

\begin{acknowledgments}
We thank G. Baym, B. L. Lev, S. Zhang and M. Lu for helpful discussions. This work was
supported in part by the National Science Foundation, under grant DMR
0758462 at the University of Illinois, and by the Office of Science, U.S.
Department of Energy, under Contract DE-FG02-07ER46453 through the Frederick
Seitz Materials Research Laboratory at the University of Illinois.
\end{acknowledgments}

\appendix

\section{The distortion of the Fermi surface of a one-component dipolar Fermi gas}
\label{app:distortion}

One of the most interesting theoretical predictions is a FS which is elliptical in shape with symmetry along the magnetization of the system. Such predictions have been made for a dipolar one component Fermi gases\cite{Miyakawa2008,Sogo2009,Fregoso2009a} and recently for two component Fermi gases\cite{Fregoso2009}. The fact that the ground state FS have elliptical shape in fully polarized systems can be understood in several ways a) by calculation of the self energy corrections to the single particle energy levels\cite{Fregoso2009a}, see appendix \ref{app:self_energy_fully_polarized} for details. b) by noting that if the density is kept constant it acts as a source field which entangles itself with the quadrupolar mode\cite{Fregoso2009a}. c) by noting that the symmetry of the bare interactions in a fully polarized system is invariant under rotations about the net magnetization axis. An elliptical FS with with same axis of symmetry is consistent with the symmetry is a possible ground state. This argument, however, does not tell whether the FS is prolate or oblate\cite{Fregoso2009a}. Here we provide perhaps the simples qualitative argument to understand the effect. It is based on elementary kinematics of a particle in an anisotropic potential and the Pauli principle.  Consider a gas uniformly polarized along $z$-axis. The Pauli principle prevents occupation of the same single particle states and hence gives some stability to the system.\cite{Fregoso2006} The interaction between two dipoles is $V(\v{r})=(\mu^2/r^3)(1- 3\cos^2\theta_{\v{r}})$, where $\theta_{\v{r}}$ is the angle of the relative position vector with the $z$-axis. As we see, the energy of particles in a region close to the $xy$-plane($\theta_{\v{r}}\sim\pi/2$) experience, on average, a repulsive interaction, i.e., they move ``up the hill''. The total energy per particle is  $\hbar^2\left\langle k^{2}_{xy}\right\rangle/2m + n \mu^2 = E$. The mean separation between particles is $n^{-1/3}$ and hence $\left\langle r^{-3} \right\rangle =n$, the particle density. However, particles in the spatial region close to the $z$-axis($\theta_{\v{r}}\sim 0$) experience on average an attraction, i.e., they move ``down the hill'' and hence the total energy per particle is $\hbar^2\left\langle  k^{2}_{z}\right\rangle/2m - 2 n \mu^2 = E$. Comparing this two expression we find for particles with \textit{same} total energy $\left\langle  k^{2}_{z}\right\rangle > \left\langle  k^{2}_{xy}\right\rangle$. Of course $\left\langle  k_{z}\right\rangle = \left\langle  k_{xy}\right\rangle=0$ because of translational invariance. This suggests that the kinetic energy and hence the Fermi wave vector of particles close to the $z$-axis is bigger than the kinetic energy of particles close to the $xy$-plane. Defining $K_{Fx} = \hbar^2\left\langle  k^{2}_{Fxy}\right\rangle/2m$ and  $K_{Fz} = \hbar^2\left\langle  k^{2}_{Fz}\right\rangle/2m$ from the relations shown above we obtain 
\beq
K_{Fz} - K_{Fx} &\sim& 3nd^2.
\eeq
As we would expect, the difference in the Fermi energies along the $z$-axis and $xy$-plane is proportional to the dipolar energy per particle.  For spin-independent forces in a translational invariant system the wave vector of particles, $\v{k}$, is a ``good'' quantum number. The FS, which concerns only with this quantum number is decoupled from the spin state of the particles. Surprisingly, dipole-dipole interaction furnish an example where this is no longer true and $\v{k}$ is entangled with the spin.

\section{Fourier transform of the dipolar interaction}
\label{app:fourier}

The Fourier transform of the dipole-dipole interaction is obtained by Fourier transforming the left hand side of the identity
\beq
- \frac{\partial^2}{\partial x_i \hspace{1pt}\partial x_j} \left( \frac{1}{|\v{x}-\v{x}'|} \right) &-& \frac{4\pi}{3}\delta_{ij} \delta(\v{x}-\v{x}') \nn \\
&&= \frac{1}{r^3} \big(\delta_{ij} - 3 \uv{r}_i \uv{r}_j \big),
\label{enq:dipole_identity}
\eeq
and using the well known results that the Fourier transform of $1/|\v{x}-\v{x}'|$ is $4\pi/q^2$. Here $\v{r} \equiv \v{x}-\v{x}'$, $r \equiv |\v{x}-\v{x}'|$ and $\uv{r}$ is a unit vector along $\v{r}$. The contact term does not contribute for finite separation between dipoles but is required because the diagonal elements( $i=j$) on the left hand side must satisfy the Poisson equation $-\nabla^2(1/|\v{x}-\v{x}'|) = 4\pi \delta(\v{x}-\v{x}')$. This makes the left hand side and right hand side traceless. It is also required because the magnetic field is divergence free. One could see this property directly from Eqn. \eqref{eqn:FT_dipole}. Note also that the dipole interaction is only valid at long distances and hence here we only consider the physics of a system of particles whose separation is larger that the Bohr radius, i.e., the electronic clouds of the atoms do not overlap.

\section{Self energy of fully polarized dipolar gas}
\label{app:self_energy_fully_polarized}
In this section we consider a homogeneous Fermi gas interacting with dipolar forces. The results of this section apply to electric as well as magnetic dipoles. The calculation below was outlined in Ref. [\onlinecite{Fregoso2009a}]. Using the formula for the decomposition of a plane wave into  a sum of spherical harmonics we can show that the Fourier transform of the dipolar interaction $V_{d}(\v{r}) = (\mu^2/r^3)(1-3 \cos^2\theta_{\v{r}})$, can be written as 

\beq
V_d(\v{k}-\v{k}') &=& \int d^3 r e^{-i (\v{k}-\v{k}')\cdot \v{r}} V_d(\v{r}) \nn \\
&=& \sum_{lm;l'm'} V_{lm;l'm'}(k,k')Y_{lm}^{*}(\uv{k}')Y_{l'm'}(\uv{k}) \nn\\
&&
\eeq
Where 
\beq
V_{lm,l'm'}(k,k') &=& -2 d^2  (4\pi)^2 (-i)^l i^{l'} \nn \\
&&\times \left(\int \frac{d r}{r} j_l(kr) j_{l'}(k'r)\right) \nn \\
&&\times \left(\int d\Omega \c Y_{lm}^{*}(\uv{r}) Y_{l'm'}(\uv{r})P_{2}(\uv{r}) \right) \nn \\
&&
\eeq
The Hartree contribution  to the self energy  vanishes because $V_d(\v{q}=0)=0$. 
The Fock term can be written as 

\beq
\Sigma_{HF}(\v{k'}) &=& -\int \frac{d^3 k}{(2\pi)^3} V(\v{k}-\v{k}') n_k^0 \nn \\
&\equiv& \sum_{lm} \sigma_{lm}(k) Y_{lm}(\uv{k}') \nn \\
&&
\eeq
Where we defined 
\beq
\sigma_{l,m}(k) &=& -\sum_{l'm'}\int \frac{d^3 k}{(2\pi)^3} V_{lm,l'm'}(k,k')Y_{l'm'}(\uv{k}) n^0_{k} \nn \\
&& 
\label{eqn:magnitude_SE_fully_pol}
\eeq
In perturbation theory we use the spherical distribution $n_k^0$ in momentum space.
We calculate the term $V_{00,lm}$ and substitute into Eqn. \eqref{eqn:magnitude_SE_fully_pol} 
to obtain
\beq
\sigma_{lm}&=& -\delta_{l,2}\delta_{m,0} \frac{4 d^2}{\pi}\sqrt{\frac{4\pi}{5}} \int_0^{k_F^0} k'^2 dk'\nn \\
&&\hspace{30pt}\times \int_0^{\infty} \frac{dr}{r} j_0(kr)\c j_2(k'r) 
\eeq
Where $k^0_F$ is the undistorted(spherical) Fermi wave vector. 
We conclude that the Hartree-Fock self energy is $\sim P_2(\uv{k})$, explicitly
\beq
\Sigma_{HF}(\v{k}) = \sigma(k)P_2(\uv{k}) 
\eeq
where $\sigma(k)$ is a smooth slowly varying function of the magnitude of the wave vector
\beq
\sigma(k) &=& \frac{-d^2}{36\pi k^3}\bigg[ 3 k^5 k_F^0 +8 k^3 (k_F^0)^3- 3 k (k_F^0)^5 \nn \\
&& - 3 (k^2 - (k_F^0)^2)^3 \arctan(\frac{k_{<}}{k_{>}}) \bigg] 
\eeq
This equation is valid for all $k$ with $k_{<}$ being the least of $k_F^0$ and $k$. One can check that
$\sigma(k)$ is continuous and twice differentiable at $k_F^0$.
We conclude that the single particle dispersion to first order is given 
\beq
\epsilon_\v{k} &=& \epsilon^0_k + \Sigma_{HF}(\v{k}) \nn \\
&=& \frac{k^2}{2m} + \sigma(k)P_2(\cos\theta_{k}) 
\eeq
In particular the Fermi wave vector can be written as 
\beq
\v{k}_F = \uv{n} k^0_{F} + \uv{n} \delta k_F 
\eeq
To first order in the dipolar interaction the 
chemical potential is unchanged, $\mu= (k^0_F)^2/2m$. The first correction to the 
chemical potential is quadratic order in the interaction. The Fermi wave vector 
is given 
\beq
\frac{\v{k}_F^2}{2m} + \sigma(k_F)P_2(\uv{k}_F) = \mu 
\eeq
Solving to first order in $\delta k_F$ we obtain
\beq
\delta k_F = \frac{m d^2 (k^0_F)^2}{9 \pi}(3 \cos^2\theta_\v{k} -1) 
\eeq
Which means that the FS is elliptical in the weak coupling limit for fully polarized dipolar gas.

\section{Self energy of an imbalanced two-component dipolar Fermi gas}
\label{app:self-energyS11andS12}

In this section we compute in Hartree-Fock the self energy of a polarized spin-\nicefrac{1}{2} system. Starting from the expression 
\beq
\hbar \Sigma_{\alpha\beta}(\v{k}) &=& -\int \frac{d^3 k'}{(2\pi)^3} \frac{1}{\hbar \beta}\sum_{n'} \mathrm{e}^{i \omega_{n'} 0^{+}}\nn \\
&&\times G^{0}_{\delta\gamma}(\v{k}',i\omega_{n'}) V_{ij}(\v{k}-\v{k}') \sigma^{i}_{\alpha\delta}\sigma^{j}_{\gamma\beta},
\eeq
one can show that $\Sigma_{21} = \Sigma_{12}^{*}$ and $\Sigma_{22}=-\Sigma_{11}$. Performing the spin indexes summation and the Matsubara sums we obtain
\begin{widetext}
\beq
\hbar \Sigma_{11}(\v{k}) &=& -\frac{4\pi\mu^2}{3} \int \frac{d^3 k'}{(2\pi)^3} (3 \cos^2\theta_{\v{k}-\v{k'}} - 1)\left[ n_F(\epsilon^{0}_{\v{k}'}-\mu_1)-n_F(\epsilon^{0}_{\v{k}'}-\mu_2)\right]
\label{eqn:sigma11}  \nn \\
\hbar \Sigma_{12}(\v{k}) &=& -\frac{4\pi\mu^2}{3}\int \frac{d^3 k'}{(2\pi)^3}  3 \cos\theta_{\v{k}-\v{k'}}\sin\theta_{\v{k}-\v{k'}}\mathrm{e}^{-i\phi_{\v{k}-\v{k'}}} 
\left[ n_F(\epsilon^{0}_{\v{k}'}-\mu_1)-n_F(\epsilon^{0}_{\v{k}'}-\mu_2)\right] \nn \\
&&
\label{eqn:self-energyS11andS12}
\eeq
\end{widetext}
Note that these expressions vanish for a system with equal number of particles in each of the 
$\sigma^{z}$ eigenstates. By expanding the dipolar interaction in spherical harmonics and using the unperturbed momentum distribution of particles , the integrals can be calculated analytically with the result
\beq
\hbar \Sigma_{11}(\v{k}) &=& -f(k) (3 \uv{k}^{2}_z  -1) 
\label{eqn:expression-sigma11} \nn \\
\hbar \Sigma_{12}(\v{k}) &=& -f(k) 3 \uv{k}_z (\uv{k}_x - i\uv{k}_y)
\label{eqn:expression-sigma12} 
\eeq
where $f(k)$ is a smooth function of the magnitude of the wave vector, 
\beq
f(k)&=& \frac{2\mu^2}{\pi} \int^{\infty}_0 {k'}^{2} dk' \int^{\infty}_{0} \frac{dr}{r} j_{2}(kr) j_{0}(k'r) \nn \\
&&\times \left[ n_{F}(\epsilon^0_\v{k'} - \mu_1) -n_{F}(\epsilon^0_\v{k'} - \mu_2) \right] 
\eeq
The self energy can be expanded in Pauli matrices as 
$\Sigma= \Sigma_0 \mathbf{1} + \Sigma_1 \sigma^x+ \Sigma_2 \sigma^y+ \Sigma_3 \sigma^z$ and we obtain
\beq
\hbar\Sigma(\v{k}) &=&  
\left(\begin{array}{cc}
 \hbar\Sigma_{11} & \hbar\Sigma_{12} \\
\hbar\Sigma^{*}_{12} & - \hbar\Sigma_{11}
\end{array}\right) = \Sigma_{i}\sigma^{i} 
\eeq
where $\Sigma_{i}(\v{k}) = -f(k)(3 \uv{k}_{i}\uv{k}_{z} -\delta_{i,z})$. We know give an explicit derivation of Eqn. \eqref{eqn:expression-sigma12}. Starting from Eqn. \eqref{eqn:self-energyS11andS12} and expanding in spherical harmonics 
\beq
V_{20}(\v{k}-\v{k}')&\equiv& \frac{4\pi\mu^2}{3}(3\cos^2\theta_{\v{k}-\v{k'}}-1)\nn \\
&=&\sum_{lm}\sum_{l'm'} V^{lm;l'm'}_{20}(k,k')Y^{*}_{lm}(\uv{k'})Y_{l'm'}(\uv{k})     \nn \\
&&
\eeq
Substituting back into Eqn. \eqref{eqn:sigma11}, we obtain the form of $\Sigma_{11}$ is 
\beq
\hbar\Sigma_{11}(\v{k})= - \sum_{l'm'} \sigma_{l'm'}(k) Y_{l'm'}(\uv{k}), 
\eeq
where $\sigma_{l'm'}(k)$,  depend only on the magnitude of the wave vector, $k$,
\beq
\sigma_{l'm'}(k)&=& \sum_{lm}\int \frac{d^3 k'}{(2\pi)^3} \c V^{lm;l'm'}_{20}(k,k')Y^{*}_{lm}(\uv{k'})\nn \\
&& \times \left[ n_{F}(\epsilon^{0}_{\v{k}'}-\mu_1) - n_{F}(\epsilon^{0}_{\v{k}'}-\mu_2)\right] 
\label{eqn:appendix-sigma_lm}
\eeq
The angular integration is simplified because the unperturbed distribution function 
is spherically symmetric. We obtain
\beq
\sigma_{l'm'}(k)&=& \frac{\sqrt{4\pi}}{(2\pi)^3}\int {k'}^{2} dk'  V^{00;l'm'}_{20}(k,k')\nn \\
&&\times \left[ n_{F}(\epsilon^{0}_{\v{k}'}-\mu_1) - n_{F}(\epsilon^{0}_{\v{k}'}-\mu_2)\right]
\label{eqn:appendix-sigma_lm2}
\eeq
The terms $V^{00;l'm'}_{20}(k,k')$  can be found by noticing that the Fourier transform of $V_{20}(\v{r}) = (\mu^2/r^3)(1-3\cos^2\theta_{\v{r}})$ is $V_{20}(\v{q})=(4\pi\mu^2/3)(3\cos^2\theta_{\v{q}}-1)$ and by definition one can write 
\beq
V_{20}(k-k')=\int dr^3 e^{-i (\v{k}-\v{k'})\cdot \v{r}} V_{20}(\v{r})
\label{eqn:V20_expansion}
\eeq 
Note that $V(\v{r})$ can also be  written as 
\beq
V_{20}(\v{r}) = -\frac{\mu^2}{r^3}\c 2\sqrt{\frac{4\pi}{5}} Y_{20}(\uv{r}) 
\eeq
and substituting the spherical wave expansion of a plane wave
\beq
e^{-i\v{k}\cdot \v{r}} = 4\pi \sum_{lm} (-i)^l j_{l}(kr) Y_{lm}(\uv{k})Y^{*}_{lm}(\uv{r}) 
\label{eqn:plane_spherical_expansion}
\eeq
into Eqn. \eqref{eqn:V20_expansion} we can read the only non-vanishing coefficient
\beq
&&V^{00;20}_{20}(k,k') = (4\pi)^2 \mu^2 \c 2\sqrt{\frac{4\pi}{5}} Y_{00} 
\int^{\infty}_{0}\frac{dr}{r} j_2(kr)j_0(k'r).   \nn \\
&&
\eeq
Substituting back into Eq. \eqref{eqn:appendix-sigma_lm2} we obtain the only non-vanishing coefficient
\beq
\sigma_{20}(k) &=& 2\sqrt{\frac{4\pi}{5}}  \frac{2}{\pi}\mu^2 \c \int {k'}^{2} dk' 
\int \frac{dr}{r}\c j_2(kr) j_0 (k'r) \nn \\
&&\times \left[ n_{F}(\epsilon^0_{\v{k'}}-\mu_1) -n_{F}(\epsilon^0_{\v{k'}}-\mu_2)\right] \nn \\
&=&  2\c \sqrt{\frac{4\pi}{5}} f(k).
\eeq
At zero temperature the unperturbed Fermi surfaces are sharply defined by wave vectors $k_{F1}=(6\pi^2 n_1)^{1/3}$ and $k_{F2}=(6\pi^2 n_2)^{1/3}$ and $f(k)$ can be analytically computed as $f(k) = g(k,k_{F1}) - g(k,k_{F2})$ where $g(x,y)$ is given by
\beq
g(x,y) &=& \frac{\mu^2}{76\pi}\frac{1}{x^3}\bigg[3x^5 y + 8 x^3 y^3 - 3x y^5 \nn \\
&&\hspace{40pt}- \frac{3}{2}(x^2-y^2)^3 \ln\left| \frac{y+x}{y-x} \right| \bigg] \nn\\
&&
\eeq
Finally, 
\beq
&&\hbar\Sigma_{11}(\v{k}) = - f(k)(3\cos^2\theta_{\v{k}}-1) = -f(k) (3 \uv{k}^{2}_z - 1) \nn\\
&&
\eeq
If $\mu_1 > \mu_2$ then $k_{F1}>k_{F2}$ and $f(k)>0$. A similar calculation yields the quoted result for $\Sigma_{12}$.


\end{document}